\documentclass[twocolumn,preprintnumbers,nofootinbib,prl]{revtex4}
\UseRawInputEncoding

\usepackage{amssymb,amsmath,graphicx}
\usepackage{epsf}
\usepackage{graphicx,epsfig}
\usepackage{amsfonts}
\usepackage{amssymb}
\usepackage{xcolor}

\usepackage{hyperref}
\hypersetup{
	unicode=false,          
	pdftoolbar=true,        
	pdfmenubar=true,        
	pdffitwindow=false,     
	pdfstartview={FitH},    
	pdftitle={My title},    
	pdfauthor={Author},     
	pdfsubject={Subject},   
	pdfcreator={Creator},   
	pdfproducer={Producer}, 
	pdfkeywords={keyword1} {key2} {key3}, 
	pdfnewwindow=true,      
	colorlinks=true,       
	linkcolor=red,          
	citecolor=cyan,        
	filecolor=magenta,      
	urlcolor=green,           
	linktocpage=true
}

\def\CL{{\cal L}}

\usepackage{soul}

\def\half{\frac{1}{2}}

\begin{document}

	\newcommand{\be}{\begin{equation}}
		\newcommand{\ee}{\end{equation}}
	\newcommand{\bea}{\begin{eqnarray}}
		\newcommand{\eea}{\end{eqnarray}}
	\newcommand{\barr}{\begin{array}}
		\newcommand{\earr}{\end{array}}
	\def\bal#1\eal{\begin{align}#1\end{align}}
	
	\pagestyle{plain}
	

	\title{Uncovering the History of Cosmic Inflation \\from Anomalies in Cosmic Microwave Background Spectra}

	\author{Matteo Braglia$^{1,2}$, Xingang Chen$^{3}$, and Dhiraj Kumar Hazra$^{4,2}$}
	\affiliation{\vspace{2mm}
		$^1$ Instituto de Fisica Teorica, Universidad Autonoma de Madrid, Madrid, 28049, Spain \\
		$^2$ INAF/OAS Bologna, via Gobetti 101, I-40129 Bologna, Italy \\
		$^3$ Institute for Theory and Computation, Harvard-Smithsonian Center for Astrophysics, 60 Garden Street, Cambridge, MA 02138, USA  \\
		$^4$ The  Institute  of  Mathematical  Sciences,  HBNI,  CIT  Campus, Chennai  600113,  India
	}

	
	\begin{abstract}
		We propose an inflationary primordial feature model that can explain both the large and small-scale anomalies in the currently measured cosmic microwave background anisotropy spectra, revealing a clip of adventurous history of the Universe during its primordial epoch. Although the model is currently statistically indistinguishable from the Standard Model, we show that planned observations such as the Simons Observatory, LiteBIRD and CMB-S4 will complement each other in distinguishing the model differences due to their accurate E-mode polarization measurements, offering very optimistic prospects for a detection or exclusion. The model predicts a signal of classical primordial standard clock, which can also be used to distinguish the inflation and alternative scenarios in a model-independent fashion.\end{abstract}
	\maketitle
	\thispagestyle{plain}
	
	{\em Introduction}.
	The inflation scenario~\cite{Starobinsky:1979ty,Guth:1980zm,Linde:1981mu,Albrecht:1982wi,Starobinsky:1980te,Sato:1980yn,Mukhanov:1981xt} is the leading candidate theory for the primordial Universe that started the Big Bang. There are high hopes that this knowledge will be advanced more definitively with future astrophysical observations, and we will be able to answer the important questions such as: Can we rule out alternative scenarios to inflation, or vice versa, using experimental data? For inflation models, can we learn any details beyond the broad-brush picture that the inflationary universe was dominated by a form of vacuum energy and expanding with acceleration?
	
	To meet these goals, experimental information beyond the Standard Model of cosmology is necessary.
	An important candidate of such information is signals of primordial features. (See~\cite{Chen:2010xka,Chluba:2015bqa,Slosar:2019gvt} for reviews.) Primordial features are strongly-scale-dependent deviations from the otherwise approximately scale-invariant spectra of the primordial density perturbations. These spectra are being probed by a variety of observations that measure the large-scale distribution of various contents of the universe.
	So far, measurements of the cosmic microwave background (CMB) provide the most precise data about the power spectrum of the density perturbations~\cite{Peiris:2003ff,Akrami:2018odb}. Although being consistent with a featureless power-law spectrum, the temperature and polarization spectra of CMB anisotropies exhibit several statistically marginal feature-like anomalies in both large and small scales~\cite{Peiris:2003ff,Akrami:2018odb}.

	At large scales, a dip in the spectrum has been noticed around $\ell\sim 20$ since WMAP~\cite{Peiris:2003ff}. At small scales, there is an oscillatory feature near $\ell\sim 750$ in the Planck data~\cite{Akrami:2018odb,Hazra:2014jwa}. Most of the model-building efforts so far have been focused on  addressing either one of these anomalies, because they have qualitatively different characters.
	
	It is well-known that the large scale anomaly may be explained by a step-like sharp feature in the inflationary potential~\cite{Peiris:2003ff,Starobinsky:1992ts,Adams:2001vc, Bean:2008na, Mortonson:2009qv, Hazra:2010ve, Hazra:2014goa, Miranda:2014fwa}, 
	in which the dip is part of the signature sinusoidal running of sharp features, and the oscillation amplitude can be made to decay quickly towards smaller scales in order to agree with the data.
	
	The small-scale anomaly has several possible explanations. It might be due to a sharp feature signal which starts at a larger scale \cite{Chen:2014cwa}. It is unlikely the extension of the previously mentioned large-scale dip, because the amplitude of the latter decays very quickly towards small scales.
	Another possible explanation~\cite{Akrami:2018odb} is the resonant feature~\cite{Chen:2008wn,Flauger:2009ab,Flauger:2010ja,Chen:2010bka,Aich:2011qv}, which can arise, for example, from periodic ripples in an inflationary potential. This explanation does not address the large-scale anomaly.
	
	On the other hand, it has been noticed in~\cite{Chen:2014joa,Chen:2014cwa} that these two anomalous features may share the same origin through the classical primordial standard clock (CPSC) effect~\cite{Chen:2011zf,Chen:2011tu,Chen:2012ja}, which we summarize below. From a top-down model-building point of view, the inflaton trajectory is determined by low-energy valleys of a potential landscape formed by many fields. It is natural to expect that such a path may not be straight and smooth (namely, have sharp features), and to expect steep cliffs perpendicular to this path (namely, the existence of many massive fields). Sharp features may temporarily disturb the inflaton away from its eventual attractor trajectory, and during the recovery process from the disturbance, some massive fields may be excited temporarily, oscillating and then settling down around potential minima after a few $e$-folds.
	In this picture, the disturbance generates the sinusoidal-running signal as a candidate for the large-scale anomaly, and the high frequency oscillation of a massive field generates the resonant-running signal as a candidate for the small-scale anomaly.
	
	Despite this interesting possibility in model building, very few explicit CPSC models have been constructed and none of them fully compared with data ~\cite{Chen:2014joa,Chen:2014cwa}. It is challenging to efficiently compare complicated feature models with data, because of the sensitivity of feature model predictions on background evolution and model parameters, and because of the multi-modal posterior distributions of feature parameters in data analysis.
		The purpose of this work is to resolve these challenges in data analyses and really start the model-searching process that requires feedback from data, making use of a recently developed methodological pipeline~\cite{Braglia:2021ckn}.
		With these advances, the sensitivity of primordial features on background evolution means that a precise measurement of these signals can tell us many details about the underlying model. Combination of the bottom-up approach, that examines the properties of the anomalies~\cite{Peiris:2003ff,Akrami:2018odb}, and the top-down approach, that classifies the phenomenological characters of different types of features~\cite{Chen:2010xka}, is often useful in putting various model ingredients into place.
	
	We will reveal a candidate model that vividly describes how, during the initial moment of the Universe, the inflaton is rolling at the top of an adventurous potential which is nonetheless quite natural from the point of view of a landscape. The emerging picture is drastically different from that of a single-field slow-roll model, but the feature signals in the CMB are all small corrections. 
		Furthermore, the resonant part of this signal (namely, the clock signal), induced by the oscillation of a massive field and taking a mostly model-independent form, directly measures the scale factor of the primordial Universe as a function of time $a(t)$ \cite{Chen:2011zf,Chen:2011tu,Chen:2012ja,Chen:2014joa,Chen:2014cwa}. Since $a(t)$ is the defining property of a primordial universe scenario, if measured it can be used to rule out alternative scenarios in a model-independent fashion.
	
	The best-fit CPSC model we find is currently statistically indistinguishable from the Standard Model. Nonetheless, this example illustrates the potential promise of this approach and the prospects of learning the history of the primordial universe from data.
	
	We further demonstrate a nice future prospect, making use of the model predictions in the polarization spectra of CMB that are strictly correlated with those in the temperature spectrum.
	There are several ongoing (e.g.~BICEP/Keck Array \cite{Moncelsi:2020ppj}) and forthcoming experiments (e.g.~Simons Observatory (SO)~\cite{Ade:2018sbj}, LiteBIRD~\cite{Hazumi:2021yqq}, CMB-S4 \cite{Abazajian:2019eic} ) in the following decade that will measure the polarization of CMB with unprecedented precisions. We forecast the prospects of the SO, LiteBIRD, and CMB-S4 
	in constraining such feature models,
	which currently remain elusive in the Planck data, and find that they will provide decisive evidence in favor of or against them.

	{\em The model}. 
	As summarized in~\cite{Chen:2011zf,Chen:2014cwa}, in general there are two simple requirements for a model to be qualified as a CPSC model. First, there should be two observable stages of inflation connected by a sharp feature. Second, the sharp feature classically excites a massive field. The most model-dependent part of the full CPSC signal is the amplitude of the sharp feature signal and its smooth connection to the clock signal. 
	We will use a step potential as part of the sharp feature. In two-field models and to connect with the oscillation of a massive field, the placement of this step in model configuration is also crucial and could lead to very different signals. 
	Our model is described by the following Lagrangian,
	\bal
	\CL =& -\half \left[ 1+ \Xi(\Theta) \sigma\right]^2 (\partial\Theta)^2
	-\half (\partial \sigma)^2
	- V(\Theta, \sigma) ~,
	\label{Eq:L_model}
	\eal
	where the potential $V(\Theta,\sigma)$ takes the following form~\cite{Footnote3},
	\bal
	\label{eq:potdip}
	&V (\Theta, \sigma) = V_{\rm inf} \Biggl\{
	1- \half C_\Theta \Theta^2 
	\\&+ C_{\sigma} \left[ 1-\exp \left( -  \frac{(\Theta-\Theta_0)^2}{\Theta_f^2}{\rm Heav}(\Theta-\Theta_0)  -\frac{\sigma^2}{\sigma_f^2} \right) \right]\Biggr\}\notag
	\eal
	and $\Xi(\Theta)= \xi~{\rm Heav}(\Theta-\Theta_0-\Theta_T)$.  We note that we set $M_{\rm pl}=c=1$ throughout this paper.

	\begin{figure}
		\includegraphics[width=0.9\columnwidth]{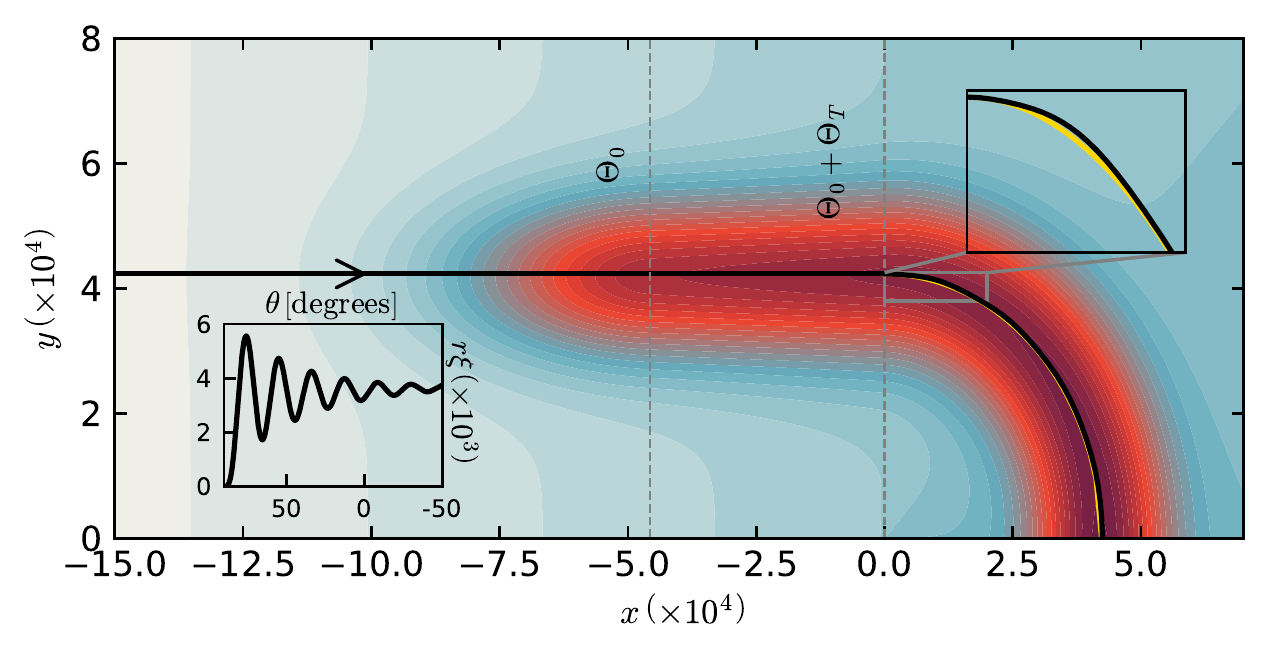}
		\caption{\label{fig:trajectory}  \footnotesize A birdseye example of background trajectory in our model, plotted over equipotential surfaces (redder means lower potential). In terms of the Cartesian coordinates $x$ and $y$ shown here, for $x<0$ the $\Theta$ and $\sigma$ are Cartesian with $x=\Theta-(\Theta_T+\Theta_0)$ and $y=\sigma+\xi^{-1}$; while for $x>0$ they become radial coordinates as in \cite{Chen:2009we,Chen:2009zp} with $r=\sigma+\xi^{-1}$ and $\theta=\pi/2-\xi(\Theta-\Theta_0-\Theta_T)$. The insets show how the inflaton overshoots the bottom of the valley (yellow line) and climbs onto a side of the cliff [top-right], and the clock oscillation [bottom-left].}
	\end{figure}
	
	This Lagrangian describes a two-field inflation model in which  $\Theta$ is the inflationary direction and $\sigma$ a field orthogonal to $\Theta$. The mathematical expressions of the potential $V$ and coupling $\Xi$ may be modified as long as they model the simple geometric configuration illustrated in Fig.~\ref{fig:trajectory}. The evolutionary history of the inflaton is described as follows.

	During the first stage of inflation, the inflaton is rolling at the top of a plateau, until it encounters a cliff, located at $\Theta=\Theta_0$ with height and width determined by $C_\sigma$ and $\Theta_f$, and falls into a lower valley~\cite{IniCond}. In the two-field space, the path at the bottom of the lower valley starts out straight and begins to curve after a distance $\Theta_T$. At the entrance of the curved path, the inflaton overshoots the bottom of the valley and climbs onto a side of the cliff, exciting the oscillation of a massive field. Due to the Hubble friction, the oscillation gradually decays and the inflaton settles down in the second stage of inflation.

	{\em Numerical results and effective parameters.}
	We follow the methodology of Ref.~\cite{Braglia:2021ckn} and directly compare numerical results on power spectrum with data.

\begin{figure}
		\includegraphics[width=.7\columnwidth]{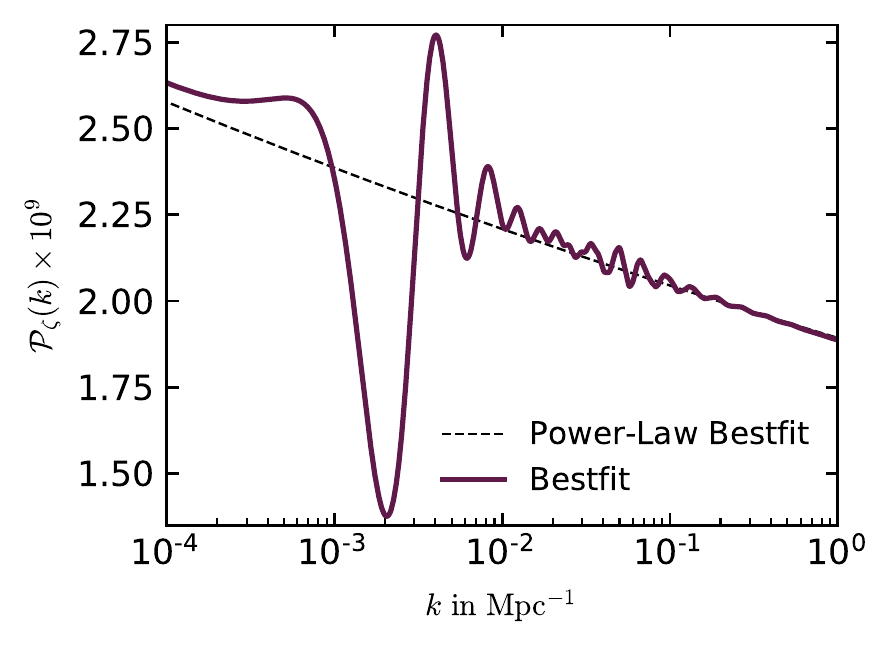}
		\caption{\label{fig:bestfit} \footnotesize The PPS of the bestfit model with
			the effective parameters: $V_{\rm inf}=5.45\times10^{-14}$, $C_\Theta=0.0189$, $C_\sigma=3.15\times 10^{-8}$, $C_\sigma/\Theta_f^2=0.495$, $m_\sigma/H=18.5$, $\xi\sigma_f=0.0580$, $N_T=1.17$, $N_0=14.38$.  These gives $\Delta P_\zeta |_{\rm dip}/P_{\zeta0}\simeq0.28$ and $\Delta P_\zeta |_{\rm clock}/P_{\zeta0}\simeq0.038$.
			In this model, $k_0 \approx 6.66\times 10^{-4} {\rm Mpc}^{-1}$, corresponding to the bottom of the dip located at $k_{\rm dip}\approx 2\times 10^{-3} {\rm Mpc}^{-1}$ or equivalently $\ell_{\rm dip}\sim20$; and $k_r\approx 4.38\times 10^{-2} {\rm Mpc}^{-1}$ or $\ell_{\rm clock} \sim 600$.}
	\end{figure}

	A typical example of the primordial power spectrum (PPS) from this model is shown in Fig.~\ref{fig:bestfit}. Note that there are two kinds of characteristically different running behaviors at large and small scales, respectively. At the large scales $k<k_r$, there is the sharp feature signal with its signature sinusoidal running $\sim \sin(2k/k_0 + {\rm phase})$, which starts near $k=k_0$. At the small scales $k>k_r$, the clock signal starts to appear with its signature resonant running $\sim ( \frac{2k}{k_r} )^{-3/2} \sin(\frac{m_\sigma}{H} \ln\frac{2k}{k_r} + {\rm phase})$. These properties are quite robust against model variations. On the other hand, the envelope of the sharp feature signal and its connection and relative amplitude to the clock signal are strongly model dependent, and play a crucial role in determining part of the model configuration.
	
	As outlined in~\cite{Braglia:2021ckn}, when comparing a multi-parameter model with data, it is often convenient to first construct an equal number of effective parameters in terms of the model parameters. Each of these effective parameters describes a distinct property of the feature signal in the power spectrum.
	The Lagrangian \eqref{Eq:L_model} contains eight model parameters. $V_{\rm inf}$ and $C_\Theta$ parameterize the scale of the inflationary potential and its slope, respectively. These two parameters are the same as those in the simplest single field model and no more transformation is needed. The other six parameters are as follows. $\Theta_0$, $C_\sigma$ and $\Theta_f$ specify the location, depth and width of the step, respectively. $\sigma_f$ describes the width of the trough, which together with some other parameters also determines the mass of the massive field $\sigma$. $\Theta_T$ gives the distance to the curved path, and $1/\xi$ is the radius of the curved path. 
	
	We first fix the slow-roll parameter $\epsilon$ to a small value, e.g.~$10^{-7}$, which determines the overall energy scale of the model. As long as $\epsilon$ is small, $\epsilon\lesssim10^{-3}$, its value does not change the phenomenology of the model after proper rescalings of other parameters. The following is the identification of the effective parameters~\cite{Braglia:2021rej}.
	
	The depth of the dip feature in the PPS is determined by the height of the step
	\bea
	\frac{\Delta P_\zeta |_{\rm dip}}{P_{\zeta0}} \approx
	1- \left( 1+\frac{3C_\sigma}{\epsilon} \right)^{-1/2} ~,
	\label{DP_dip}
	\eea
	and the extensiveness of the sinusoidal running is determined by the width of the step,
	$\Delta k \sim k_0\sqrt{C_\sigma/\Theta_f^2}$.
	These relations suggest two effective parameters: $\frac{\Delta P_\zeta |_{\rm dip}}{P_{\zeta0}}$ and $C_\sigma/\Theta_f^2$.
	
	The amplitude of the clock signal, in addition to its running property mentioned previously, is determined by the initial velocity of the inflaton entering the curved path, the radius of the path and the mass of the $\sigma$-field:
	\bea
	\frac{\Delta P_\zeta |_{\rm clock}}{P_{\zeta0}} \Big|_{\rm amp} \approx
	\frac{\sqrt{2\pi}}{3} \frac{\epsilon}{C_\sigma} (\xi \sigma_f )^2 \left( \frac{m_\sigma}{H} \right)^{1/2} ~,
	\label{DP_amp}
	\eea
	where $m_\sigma/H \approx \sqrt{6 C_\sigma}/\sigma_f$.
	This suggests two more effective parameters, $\log_{10} m_\sigma/H$ and $\frac{\Delta P_\zeta |_{\rm clock}}{P_{\zeta0}}$.
	
	The transition between the step and the curved path is most conveniently described by the number of $e$-folds the inflaton spends in-between, $N_T \approx (\Theta_T/\sigma_f) (H/m_\sigma)$, which we use as another effective parameter.
	
	The last effective parameter is the overall $k$-location of the feature, which we parameterize as $N_0$, defined as the $e$-fold from the beginning of inflation, at which the inflaton crosses the location of the step $\Theta(N_0)\equiv\Theta_0$~\cite{pivotscale}.
	
	In total, the power spectrum is specified by six more than that in the Standard Model.
	In term of effective parameters, the starting locations of sharp feature and clock signal, $k_0$ and $k_r$, are related by $k_r \approx \exp(N_T)\frac{m_\sigma}{H} k_0$.

	{\em Data comparison and best fit.}
	Our data analysis is based on publicly available latest Planck data of CMB temperature and E-mode polarization.  Following the Planck inflation paper~\cite{Akrami:2018odb}, we use $\tt{commander\_dx12\_v3\_2\_29}$ for temperature anisotropies and $\tt{simall\_100x143\_offlike5\_EE\_Aplanck\_B}$ for E-modes, both at $\ell=2-29$; at high-$\ell$, we use the unbinned {\tt Plik bin1} likelihood for TT, TE, EE anisotropies~\cite{ACT}. We refer to this dataset as P18.
	
	Multi-modality of the posterior distributions of the feature parameters is expected~\cite{Braglia:2021ckn}. Therefore we use nested sampling through the {\tt PolyChord}~\cite{PolyChord,Handley:2015fda,Handley:2015vkr} implementation in CosmoMC~\cite{Lewis:2002ah}. From the samples in the data analysis, we plot the posterior distributions of the parameters and, importantly, compute the Bayesian evidence $Z_i$ of the model $i$. The latter helps us compare our model to the featureless, baseline model.

	Together with the cosmological parameters, i.e. $\omega_b,\, \omega_c,\, \tau_{\rm reio}$ and $100*\theta_s$, we vary the inflationary parameters as well as the nuisance parameters. The priors for the effective parameters are chosen as follows: $\log_{10}P_{\zeta *}\in[-8.82,\,-8.1]$, $C_\Theta\in[0.002,\,0.04]$, $N_0\in[13,\,15.5]$, $\frac{\Delta P_\zeta |_{\rm dip}}{P_{\zeta0}}\in[0,\,0.5]$, $C_\sigma/\Theta_f^2\in[0,\,2.5]$, $N_T \in [0,\,1.2]$, $\log_{10} m_\sigma/H\in[0,\,\log_{10} 75]$, $\frac{\Delta P_\zeta |_{\rm clock}}{P_{\zeta0}}\in[0,\, 0.35]$. With these choices, the locations of both the dip feature and the clock signal are allowed to appear in the whole range of multipoles probed by Planck.
	
	\begin{figure}
		\includegraphics[width=.8\columnwidth]{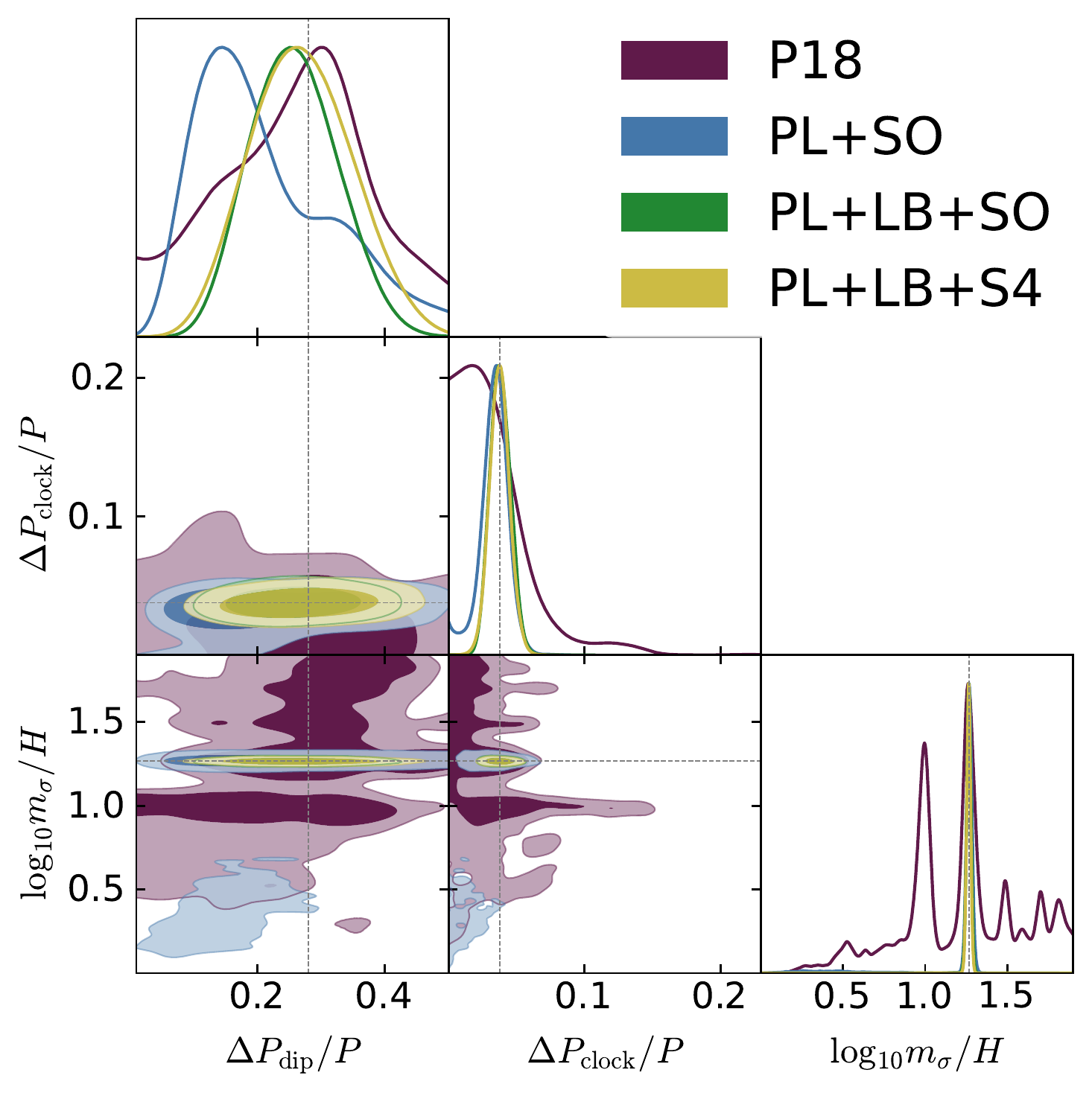}
		\caption{\label{fig:triangle} \footnotesize Constraints from P18 and projected constraints on the parameters characterizing the amplitude of the dip and the amplitude and frequency of the clock signal.}
	\end{figure}

	Planck constraints on the parameters describing the feature amplitudes and the frequency of the clock signal are shown in 
	purple in Fig.~\ref{fig:triangle}. Full triangle plots showing constraints on all parameters
	are presented in a companion paper~\cite{Braglia:2021rej}. Although our CPSC model provides a better fit than the baseline model, due to introduction of extra parameters, it is indistinguishable from the Standard Model  according to the Jeffreys' scale~\cite{Kass:1995loi} with the Bayes factor being $\ln B\equiv \ln( Z_{\rm feature}/Z_{\rm featureless})=-0.13\pm0.38$. This confirms
		no evidence of features in Planck data, consistent with previous analysis~\cite{Akrami:2018odb,Braglia:2021ckn}.
	
\begin{figure}
		\includegraphics[width=\columnwidth]{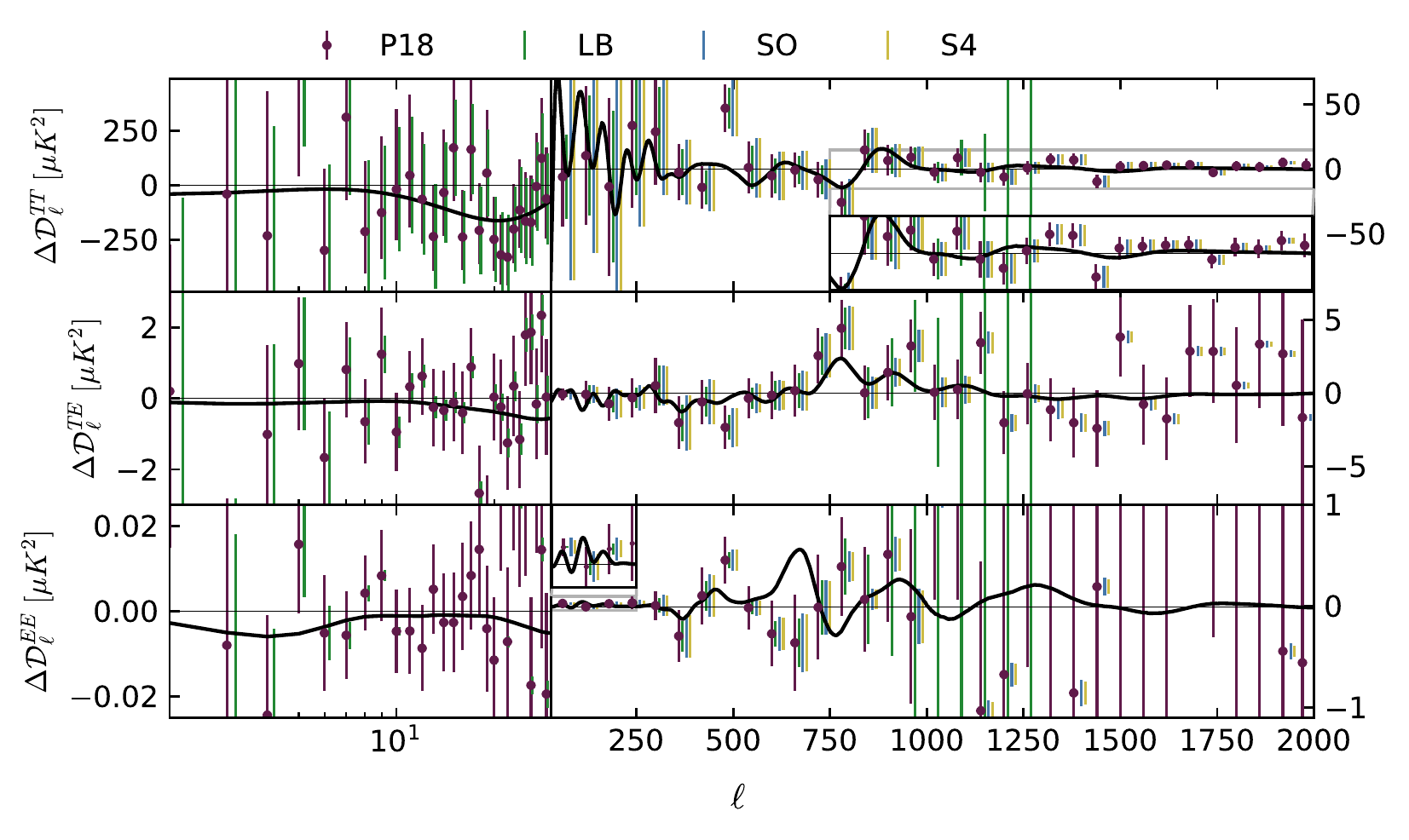}
		\caption{\label{fig:errors} \footnotesize Residual plots for our best fit candidate. P18 data points and error bars are plotted in purple, whereas the forecast error bars for LiteBIRD, SO and CMB-S4
			are plotted in 
			green, blue and gold, respectively. Errors for $\ell>30$ are binned with binwidth $\Delta\ell=60$. }
	\end{figure}

	As mentioned previously, these analyses nonetheless pick up an interesting bestfit candidate.
	A clear peak around $m_\sigma/H\sim18$ stands out in the posterior. From the analysis of the samples we observe that almost all the better likelihood points concentrate around that mode. Using BOBYQA~\cite{BOBYQA}, we obtain the bestfit candidate and quantify the improvement in the fit to P18~\cite{Footnote2}. Our bestfit candidate's PPS is presented in Fig.~\ref{fig:bestfit} and the corresponding CMB residuals in Fig.~\ref{fig:errors}. 
	The $\Delta\chi^2=19.8$ improvement over the featureless model includes those from the dip feature, fitting  low-$\ell$ data ($\Delta\chi^2_{\rm low-T}=5.34$ and $\Delta\chi^2_{\rm low-E}=1.11$), and those from the clock signal, fitting  high-$\ell$ data ($\Delta\chi^2_{{\rm high-}\ell}=13.31$).
	
	The improvement in fit can be appreciated by looking at the residual plot in Fig.~\ref{fig:errors}, which clearly shows that it is driven not only by the better fit to the dip in the TT residuals around $\ell\sim20$, but also by the sinusoidal running of the sharp feature signal and, in particular, the clock signal. The latter starts around $\ell\gtrsim 600$ and addresses the dip followed by a bump in the TT residuals around $\ell\sim750$ and the associated feature in the TE residuals.
	The sinusoidal running part of the sharp feature signal also provides some weaker improvement in fits to the bump in the EE residuals around $\ell \sim 60$ and the dip around $\ell\sim350$.

	{\em Forecast with CMB polarization.}
	Assuming the ${\cal D}_\ell^{\rm TT}, {\cal D}_\ell^{\rm EE}, {\cal D}_\ell^{\rm TE}$ from the best fit as the fiducial angular power spectra, using the projected noise power spectra from the upcoming SO, LiteBIRD,  and CMB-S4
	, we forecast future constraints on our model. SO is a ground based observation with a polarization noise an order of magnitude lower than Planck (which was $\sim 52 \mu K$). It will start taking data in 2023. LiteBIRD, planned to be launched around 2029, is a full sky CMB mission with a sensitivity of $2.2\mu K-{\rm arcmin}$ in polarization and $0.5^\circ$ resolution. While SO covers 40\% of the sky (compared to 70\% by LiteBIRD), it has a much finer resolution (better than $3'$). CMB-S4 is designed to reach a sensitivity of $1\mu K-{\rm arcmin}$ with nearly 500,000 detectors and is expected to detect the high-$\ell$ CMB peaks to a much better accuracy compared SO.  While it has not been funded yet, CMB-S4 has received strong support by the ASTRO2020 report \cite{Astro2020}. 
		The projected error-bars for the three observations are plotted in Fig.~\ref{fig:errors}.In the forecast analysis, we consider the following combinations of experiments: Planck+SO, Planck+SO+LiteBIRD and Planck+S4+LiteBIRD. We refer the readers to Ref.~\cite{Braglia:2021rej} for a detailed discussion of our analysis.
	We plot projected constraints on top of Planck ones in Fig.~\ref{fig:triangle}.

	Future experiments will play complementary roles in constraining the model. SO, while impressively increasing the constraints on the clock frequency, will not be able to definitively detect neither the clock signal nor the dip feature because of residual degeneracies induced by the poor constraints on the latter. With the launch of LiteBIRD, though, the constraining power at large scales will increase drastically, being capable of detecting the dip feature amplitude, which will be constrained away from $0$ at more than $4\sigma$. Pinning down the dip feature will also help to fully exploit the exquisite power of SO at high-$\ell$, leading to a $5\sigma$ detection forecast of the clock signal.  For Planck+LiteBIRD+SO, we find a projected Bayes factor of $+22.7$, suggesting that in less than a decade we may be able to provide decisive evidence in favor of or against our model. CMB-S4 will further increase constraints on the clock frequency.

	{\em Conclusions and discussions.}
	CMB anomalies may hint at primordial physics beyond the standard model of cosmology. In this letter, we have proposed a full classical standard clock model of inflation where a sharp feature exciting massive-field oscillations addresses the low and mid-$\ell$ anomalies, whereas anomalies at high-$\ell$ are instead fitted by the clock signal.
	The improvement in the $\chi^2$ for the global best-fit candidate, characterized by a clock field with an effective mass $\sim18$ times the Hubble scale of inflation, is $\Delta\chi^2\sim19.8$. 
	According to the Bayesian evidence, this model is currently indistinguishable from the Standard Model.
	Assuming such a candidate as a fiducial cosmology, we have performed a forecast for future CMB experiments and highlighted the complementarity of measurements of E-mode spectra across different scales. We find these experiments offer promising prospects within the next decade: Simons Observatory and LiteBIRD, joint with the Planck data, will be able to place significant constraints on all parameters of our model, and CMB-S4 will further improve these constraints. These results also suggest promising prospects of model-building and testing of primordial feature models such as the one presented in this work. If detected, such a model can provide vital information about the origin of the Big Bang, ranging from a direct evidence for the inflation or an alternative scenario to detailed dynamics of the inflation model.
	
	Besides signatures in CMB, primordial feature models also leave correlated imprints in the large-scale distributions of galaxies~\cite{Huang:2012mr,Chen:2016vvw,Ballardini:2016hpi,Palma:2017wxu,LHuillier:2017lgm,Ballardini:2017qwq,Beutler:2019ojk,Ballardini:2019tuc,Debono:2020emh,Li:2021jvz} and atomic hydrogen~\cite{Chen:2016zuu,Xu:2016kwz}, which will be further tested in future Large-Scale Structure and 21cm observations. Besides power spectrum, feature models also generate correlated signals in primordial non-Gaussianities~\cite{Chen:2006xjb,Chen:2008wn,Adshead:2011jq,Hazra:2012yn, Bartolo:2013exa,Fergusson:2014hya,Fergusson:2014tza}. It would be interesting to compute the bispectrum of this model and examine its observability.
	
	As mentioned, the anomalies in the CMB power spectra can also be fit~\cite{Akrami:2018odb} by other inflationary models, including pure sharp feature models~\cite{Starobinsky:1992ts,Adams:2001vc,Chen:2006xjb,Bean:2008na, Hazra:2010ve,Achucarro:2010da,Gao:2012uq, Hazra:2014goa,Canas-Herrera:2020mme,Miranda:2014fwa} and simple resonant models~\cite{Chen:2008wn,Flauger:2009ab,Flauger:2010ja,Chen:2010bka,Aich:2011qv}. It is also possible that such features are generated by models of alternative scenarios to inflation~\cite{Chen:2018cgg,Domenech:2020qay}, or models containing non-standard primordial clocks~\cite{Chen:2011zf,Huang:2016quc,Domenech:2018bnf,Wang:2020aqc}. Our forecast on the size of error bars from future CMB polarization experiments offers some optimistic notes on the prospects of experimentally distinguishing many of these different cases if the amplitude of the signal is similar to that of the best-fit model in this Letter, although these aspects deserve to be studied more extensively.
	
	All these add to the exciting prospects of probing the history of the primordial Universe using data from near future observations.

	{\em Acknowledgment.}
	We thank Mario Ballardini, Karim Benabed, Josquin Errard, Steven Gratton, John Kovac, Umberto Natale, Luca Pagano, Daniela Paoletti and Daniela Saadeh for very helpful discussions.
	Computations were run on the FASRC Cannon cluster at Harvard University. MB is supported by the Atracci\'{o}n de Talento contract no. 2019-T1/TIC-13177 granted by the Comunidad de Madrid in Spain.
	

\begin{thebibliography}{10}
\bibitem{Footnote3}
		In order not to make confusion with the field $\Theta$ we denote the Heaviside Theta function by ${\rm Heav}$. Throughout this paper we use the convention $M_{\rm Pl}=1$.
\bibitem{pivotscale}
		We choose $k=0.05\,{\rm Mpc}^{-1}$  for the mode that exits the horizon $50$ $e$-folds towards the end of the inflation. We fix a total of $68$ $e$-folds.
\end{thebibliography}

\end{document}